\documentstyle[editedvolume,numreferences,psfig]{kluwer_astroph}
\newcommand{\stt}{\small\tt}
\begin{opening}

\title{The Selection\protect\\
       of Tenured Astronomers in France}

\author{G.A. Mamon}
\institute{Institut d'Astrophysique\footnote{CNRS UMR 7095}\\
           98 bis, boulevard Arago\\
           F-75014 Paris, France \\
           \stt gam@iap.fr}

\end{opening}
\runningtitle{Selection of Astronomers in France}

\begin{document}

\begin{abstract}
The organization of the recruitment of tenured astronomers and
astrophysicists in France is presented and compared with the setup in other
countries. 
The ages for getting tenure have increased from 27-28 in 1980 
to 31 today.
Foreign scientists constitute at least 11\% of the recruits and the delay 
in their hiring is quantified.
The large reliance on national tenure committees is justified, while
the increased targeting of positions is questioned and a compromise
proposed.
\end{abstract}

\section{Introduction}

The international reputation of various nations in a particular scientific
domain is not simply proportional to the total money spent on it.
Among other factors, one could list: the organizational setup, the focus on
the field
by the national instances, the spirit of competition between different
laboratories. But perhaps the single most important factor leading to the
success of a nation in a given scientific field is how it trains and
selects its scientists.

In this article, I will describe how the tenured scientists are selected in
the fields of astronomy and astrophysics in France.
I will limit myself to scientists, and not discuss the recruitment of
engineers, technicians and administrative staff.
By astronomy and astrophysics, I will span the fields of solar physics,
space physics (solar wind and magnetospheres including terrestrial),
planetary physics, stellar
physics, high-energy astrophysics, interstellar and circumstellar physics and
chemistry, galactic
structure and extragalactic science, cosmology (from the local universe to
the cosmic microwave background), and physics ({\em e.g.} calibration of
atomic and molecular parameters of astronomical relevance).

I will describe the three major corps of scientists in Sec.\ 2 and the
selection committees and procedures in Sec.\ 3, illustrate
the evolution of the hiring in
Sec.\ 4, compare with other countries in Sec.\ 5, and finally
list and attempt to
answer the outstanding questions in Sec.\ 6. 
A summary is provided in Sec.\ 7.

\section{The Three Major Corps}

The French scientists in the fields of astronomy and astrophysics are
subdivided, in roughly equal numbers, among three separate corps listed in
Table \ref{corps3}: the {\em Centre National de la Recherche Scientifique}
(hereafter, CNRS), the Observatories and the Universities.

\begin{table}[ht]
\renewcommand{\tabcolsep}{2pt}
\begin{center}
\caption{The three major corps of astronomers and astrophysicists in France}
\begin{tabular}{lccc}
\hline
{\bf Corps} & {\bf CNRS} & {\bf Observatories} & {\bf Universities}\\
\hline
\vspace{0.5\baselineskip}
{\bf Hiring} & CNRS & CNAP$^a$ & CNU$^b$ \\
{\bf agencies} & (Section 14)$^c$&  (Astronomy Section)$^d$ & (Section
34)$^e$+local \\
\vspace{2.4mm}
        & [national] & [national] & [mainly local] \\
{\bf Title of} & Charg\'e & Astronome-Adjoint & Ma\^{\i}tre de \\
\vspace{2.4mm}
{\bf entry-level} & de Recherche & & Conf\'erences \\
{\bf Title of} & Directeur & Astronome & Professeur \\
\vspace{2.4mm}
{\bf upper-level} & de Recherche &  & \\
\vspace{2.4mm}
{\bf Total number} & 320 & 230 & 145$^f$ \\
{\bf Number at} & 171 & 116 & 85$^f$ \\
\vspace{2.4mm}
{\bf entry level} & & & \\
{\bf Number at} & 149 & 114 & 60$^f$ \\
\vspace{2.4mm}
{\bf upper level} & & & \\
\vspace{2.4mm}
{\bf Annual hirings}$^g$ & 8.2$^h$ & 7.0 & 5.6$^h$ \\
{\bf Fraction of} & 28\% & 0 & 60\% \\
\vspace{2.4mm}
{\bf targeted positions}$^g$ & & & \\
{\bf Service} & -- & observational$^i$ & -- \\
\vspace{2.4mm}
        &       & (large projects) &  \\
\vspace{2.4mm}
{\bf Teaching load}$^j$ & 0 & 64 & 192 \\
{\bf Median age$^g$ for} & 45 & 42 & 39 \\
\vspace{2.4mm}
{\bf promotion to upper-level} \\
{\bf Mobility} & very easy & easy & difficult \\
\hline
\end{tabular}

$^a$ {\em Conseil National des Astronomes et des Physiciens} \\
$^b$ {\em Conseil National des Universit\'es} \\
$^c$ {\em Solar System and Distant Universe}
({\tt http://dasgal.obspm.fr/\~{}section/}) \\
$^d$ ({\tt http://wwwusr.obspm.fr/commissions/cnap/CNAP.html})\\
$^e$ {\em Astronomy, Astrophysics} ({\tt http://www-obs.cnrs-mrs.fr/cnu/})\\
$^f$ four Ma\^{\i}tres de Conf\'erences and four Professeurs\\
managed by Section 34 of CNU are outside the field \\
$^g$ last five years\\
$^h$ another 0.2 hired by CNRS and 1.0 hired by the Universities
 are outside the field \\
$^i$ roughly 1/3 time \\
$^j$ in equivalent hours of recitations per year
\label{corps3}
\end{center}
\end{table}

Among the approximate total of 750 French tenured astronomers and
astrophysicists, 
44\% are CNRS scientists, 30\% have been hired by the Observatories, 19\% are
University faculty, and another 7\% were hired by other institutions (almost
all at
the {\em 
Commissariat \`a l'Energie Atomique}, hereafter 
{\em CEA}, which has its {\em Service
d'Astrophysique}). 
Table \ref{corps3}
illustrates the similarities and differences among the three major corps:
roughly equal sizes, national tenure committees for two (the
CNRS and the CNAP for 
Observatory positions) of the three corps,
but different attributes, and varying difficulties for
promotion and mobility.

Note that their are two entry levels for CNRS positions ({\em Charg\'e
de Recherche} 2nd class and 1st class -- CR2 and CR1 respectively), 
and similarly two upper
levels ({\em Directeur de Recherche} -- DR2 and DR1).
Whereas two upper levels exist both for the Observatories and the
Universities, the rule of law only allows hiring to 2nd class levels.
Moreover, the Universities and Observatories have both merged their two entry
level classes.
Finally, all three corps have exceptional-class upper levels, for scientists
with exceptional careers, and the Observatories and Universities have
a higher paying entry-level for older scientists.

\section{The Selection Process}

\subsection{The tenure committees}

The members of 
French national tenure committees (CNRS and CNAP)
are mostly elected with a minority (1/4 to
1/3) nominated by high-level administrators.
The elections are usually based upon lists of candidates (as required by rule
of law), 
where for the past 10 years, the tradition has been to have typically three 
lists affiliated with national labor unions, and one independent
list. Many of the candidates on labor union lists are not personally
affiliated with these labor unions, and it appears to this author (who served
at different times on all three tenure committees) that there
is little collusion between members of a given list (unionized or not) on
important votes. The specificity of unionized lists is that they can rely on
advice from active union people that are either not on the
committees or in other scientific fields, especially for establishing their
platform and for negotiating with the government on proposed reforms.

The CNRS, CNAP and CNU committees also handle
promotions within the different levels, and the CNRS committee is also
involved in formulating policy and strategic planning, 
as well as in the analysis of
entire laboratories and national subfield networks ({\em Groupements de
Recherche}).

With the exception of the CNRS, junior scientists do not participate
in the recruitment of upper-level scientists, while all upper-level
committee members participate in the promotion of the sparse exceptional
classes.

\subsection{CNRS and Observatory positions: filled by national committees}

\parskip=0pt
The CNRS and Observatory positions are filled in a similar fashion.
All candidates are assigned two referees from the tenure committee, who
study their CV in detail, and the first referee
provides a few minute report to the
committee, while the second referee adds information in a much shorter
intervention.
All entry-level candidates pass a short audition 
(20 min + 5 min questions at the
CNRS; 15 min + 3 min questions at CNAP), as well as the candidates for
promotion to the upper level at the Observatories (the CNRS astronomy section
stopped auditioning its upper level candidates in 2001).
The very large number of candidates (of order 100) forces the committees to
split up into two to three auditioning sections, and an unsuccessful candidate
will be generally rotated to a different auditioning subsection on his/her new
attempt in the following year.
The committees then discuss and vote to narrow down the list of candidates
from typically over 100 to the final number typically less than 10.

The principal selection criterion is  the excellence of the scientific
research as estimated by the publication record, 
the analysis by the referees, 
and the audition.
There is also an increased use of citation statistics,
although there is fairly little correlation between the chosen
candidates and their publication and/or citation rates, even normalized to the
number of authors and rank of the candidate among the authors
(in part because the committees are very careful not to discriminate against
scientists who publish less, such as 
instrumentalists and people working in pure theory).

Additional criteria are the pertinence of the candidate's
research topics, his/her autonomy, ability to work in teams, perceived
usefulness to the laboratory (s)he suggests to work at,
general dynamism, student supervision, 
teaching ability for the CNAP (see Sect.\ 3.3),
educational background (being an alumnus of one of the French {\em Grandes
\'Ecoles} is a plus, and a detail rarely forgotten by the referees), and
possible awards and distinctions.
Moreover, the CNRS and CNAP often strive to maintain a balance between the
different laboratories throughout the country, as well as between the
subfields of the candidates.
In some instances, two scientists are hired in a given subfield and none in
another subfield, while in other cases, the committee will build its short
list by requiring at most one candidate per subfield (there are usually fewer
positions than subfields).

The selection of Observatory scientists is more complex than that for their
CNRS counterparts, as, in addition to the standard requirement of scientific
excellence, a successful candidate must imperatively be
attached to an observational service task (simulation of observations through
the instrument; design, assembly, or
calibration of the instrument; systematic 
observations; pipeline development; data
archival), and this service task must be for one of the dozen or so 
observational
projects that is considered top priority by the astronomical funding agency
({\em Institut National des Sciences de l'Univers} or {\em INSU}). 
For example, a brilliant theoretician without any proven track record for
the service task he proposes stands no chance of entering the corps of
Observatories. 
Unfortunately, the CNAP has little leverage on Observatory scientists who do
not perform their service tasks, given the recent merger of the two
entry-level classes.

The
discussions are tricky as committee members refrain from being negative on
candidates.  Some committees discuss more, while others resort more often to
votes.
In a blocked situation, one often displays a {\em mute
histogram}, 
without names attached (except for the person counting the votes), and one
can then
place a new rejection threshold at some local minimum in this histogram.
The advantage of discussing more is that more useful things are said, and
decisions can be made almost by consensus.
The disadvantage of too much discussion is that a few
charismatic and forceful committee members have an undue influence on the
other committee members.

The CNRS hiring committee provides a list that can be modified by a
high-level administrator at the CNRS (such changes are extremely rare), while
the list provided by the CNAP is final.

\subsection{University hiring: national qualification then local hiring}

The University positions are filled in a different fashion.
The positions are all targeted to specific subfields of astronomy and
astrophysics, although roughly 1/4 are as vague as {\em 
Astrophysics}, while roughly half are fairly narrowly targeted.
The committees are local to their universities, with roughly one-quarter of
the members nominated from outside institutions.
But these local committees can only consider candidates that have been
previously {\em qualified} by a section (usually Astronomy) of
the CNU.

There are three criteria for qualification: research activity, teaching
activity, and proficiency in the French language.
Roughly 2/3 to 3/4 of all entry-level candidates are qualified by the CNU.
The qualification is good for four years.
Less than half of the candidates for qualification to Professorship are
qualified by the CNU (given the high fraction of foreign candidates, many
of whom are not proficient in the French language).

The local tenure committees then make a short list
of typically 5 candidates or less, and audition them for
typically a half-hour, and then discuss and vote to select a candidate.
They emphasize the scientific record, 
as well as the teaching ability of the candidate as judged by recommendation
letters and the clarity of expression (the latter is also used, albeit to
a lesser degree, by the CNAP).

The chances for promotion of a Ma\^{\i}tre de Conf\'erences
to Professeur
are randomized by the requirement that a position of Professeur opens up at
his/her University, with a description that meets his/her work.

\section{Evolution Since 1980}

\subsection{Hirings per section}

Fig.\ \ref{evolnum} shows the evolution of annual hirings in the different
corps of scientists.
Since 1989, the rate of hiring has been higher than in previous years, thanks
mainly to increased hirings at the Universities, the CEA, and to some extent
at the Observatories, to reach a total annual rate of 22 hirings per
year.
Part of this increase is related to the rise of the field of astroparticles.

\subsection{Pressure}

\begin{figure}
\centerline{\psfig{file=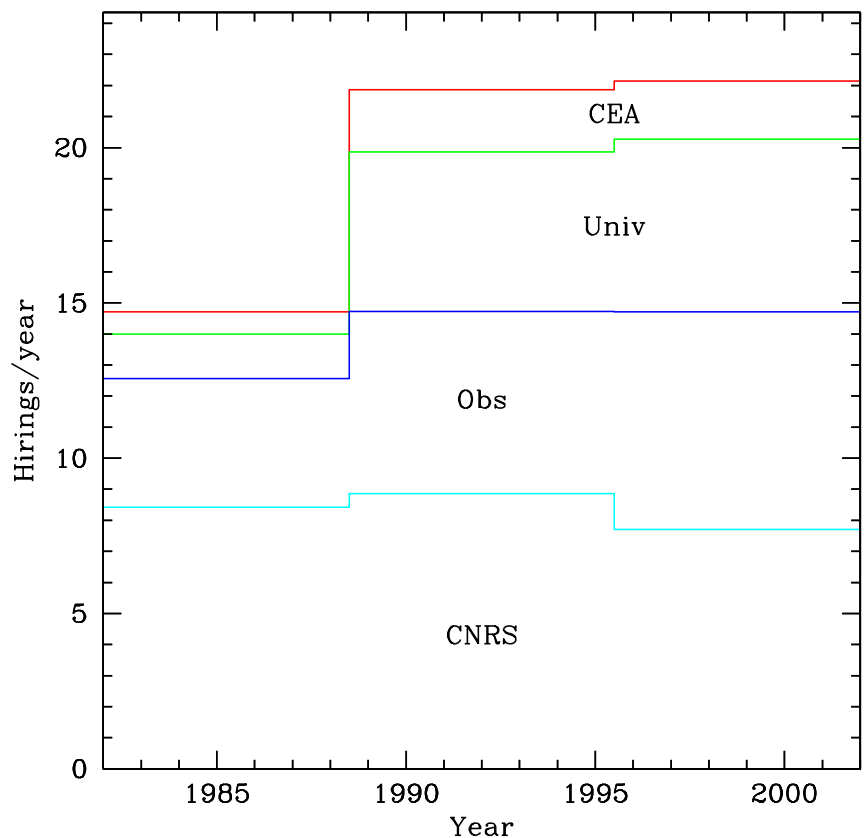,height=12cm}}
\caption{Evolution of the annual hirings in the different corps, between the
7-year periods 1982-1988, 1989-1995, 1996-2002.}
\label{evolnum}
\end{figure}

\parskip=0pt
The first half of the 90's saw a greater than doubling of the number of
candidates for tenured positions (from 35-40 to 80),
associated with only a 50\% increase in the
total number of positions (see Fig.\ \ref{evolnum}).
This dramatic rise in the number of candidates is believed to be associated
to the time when scientists with French doctorates began pursuing their
careers with postdoctoral fellowships 
in other countries, and with the slight rise in the
median age of hiring (Sec.\ \ref{ageevol} and Fig.\ \ref{evolage} below).

Since the mid-90's, the number of candidates for Observatory positions has only
risen slightly ($<10\%$) for observatory positions.
Indeed, at this time, the CNAP began to place strong emphasis on
observational service tasks as a necessary (but insufficient) selection 
criterion, so that many candidates (often theoreticians) with no obvious
service tasks, have censored themselves with respect to Observatory positions.
On the other hand, the number of
candidates to CNRS positions has continued to rise
since the mid 90's, to typically 100 today.
Given the increase in pressure on tenured positions that occurred in the
early 90's, the graduate schools in France began, in the late 90s, to reverse
their policy of increasing the number of graduate students.

It is interesting to follow the candidates to a given corps for
a given year.
Among the 37 candidates to Observatory positions in 1989, 16 (43\%) were
hired that year in one of the four corps, 8 (22\%) the following year, and
another 7 (19\%) in the subsequent 3 years, so that within 4 years, all but 5
(14\%) had found tenured scientific positions.
But, by 1994, only 50\% of the 82 candidates to the Observatories were
eventually hired in one of the four corps.

\begin{table}[ht]
\renewcommand{\tabcolsep}{5pt}
\begin{center}
\caption{Pressure on recent national positions}
\begin{tabular}{lcclcr@{\ \ \ \ }r@{\ \ \ }r@{\ \ \ }}
\hline
Year & Corps & Level$^a$ & Target$^b$ & Sites$^c$ & \multicolumn{1}{c}{Cands$^d$} & \multicolumn{1}{c}{Pos$^e$} & \multicolumn{1}{c}{Press$^f$} \\
\hline
2000 & CNRS & CR2 & -- & -- & 105 & 2 & 52 \\
2000 & CNRS & CR1 & -- & -- & 53 & 2 & 26 \\
2000 & CNRS & CR2 & IRAM & -- & 17 & 1 & 17 \\
2000 & CNRS & CR1 & AUGER & 1 & 9 & 1 & 9 \\
2001 & CNRS & CR2 & -- & -- & 96 & 4 & 24 \\
2001 & CNRS & CR1 & -- & -- & 41 & 2 & 20 \\
2001 & CNRS & CR2 & Solar system \& exoplanets & -- & 6 & 1 & 6 \\
2001 & CNRS & CR2 & Sun-Earth relations & -- & 5 & 1 & 5 \\
2001 & CNRS & CR2 & Cosmology & -- & 5 & 1 & 5 \\
2001 & CNRS & CR2 & Laboratory astrochemistry & 1 & 1 & 1 & 1 \\
2002 & CNRS & CR2 & -- & -- & 88 & 3 & 29 \\
2002 & CNRS & CR1 & -- & -- & 31 & 2 & 15 \\
2002 & CNRS & CR2 & High energy & -- & 12 & 1 & 12 \\
2002 & CNRS & CR2 & Exoplanet instrumentation & -- & 9 & 1 & 9 \\
2002 & CNRS & CR2 & Nano particles and dust & 3 & 7 & 1 & 7 \\
2001 & Obs & AA2 & -- & -- & 81 & 6 & 13 \\
2002 & Obs & AA$^g$ & -- & -- & 88 & 8 & 11 \\
\hline
\end{tabular}
\end{center}
$^a$ CR = Charg\'e de Recherche, AA = Astronome-Adjoint,
while the suffix is the class;
$^b$ abbreviated;
$^c$ number of laboratories where the position is targeted;
$^d$ number of candidates;
$^e$ number of positions;
$^f$ pressure;
$^g$ after the merger of the two AA classes.
\label{pressuretb}
\end{table}

The pressure on positions is quantified in Table 2.
The median pressure on targeted positions is 9, whereas the median pressure
is 25 on non-targeted CNRS positions and 12 on Observatory 
positions (double the
pressure on Observatory positions during the late '80s).
In 1996 and 1997, the median pressure on University positions was 23.

\subsection{Hirings per field}

\begin{figure}
\centerline{\psfig{file=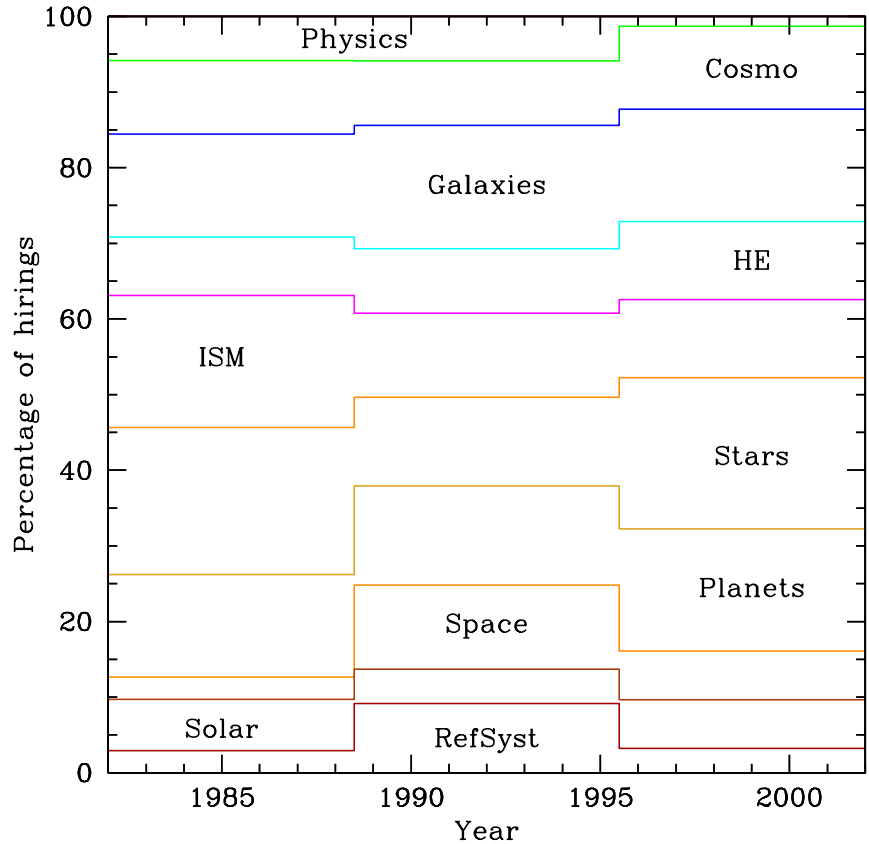,height=12cm}}
\caption{Time evolution of the fields of the new recruits in France.
>From {\em bottom} to {\em top}, the fields are (see Sec.\ 1):
Reference Systems, Solar, Space Physics, Planetary, Stellar,
Interstellar, High Energy, Galaxies, Cosmology and Basic Astrophysics (mainly
atomic and molecular).
}
\label{evolfields}
\end{figure}

Fig.\ \ref{evolfields} shows the variations with time of
the fields of the new
recruits.
The fluctuations in the fraction of hiring per field do not appear to be
statistically significant, except for the field of high-energy astrophysics,
albeit marginally so
(the probability, from
binomial statistics, of having as few High Energy positions in the first
period combined with as many High Energy positions in the third period,
assuming the actual mean rate of 9.0\% over the three periods, is
5\%).

\subsection{Hirings per methodology}

\begin{figure}
\centerline{\psfig{file=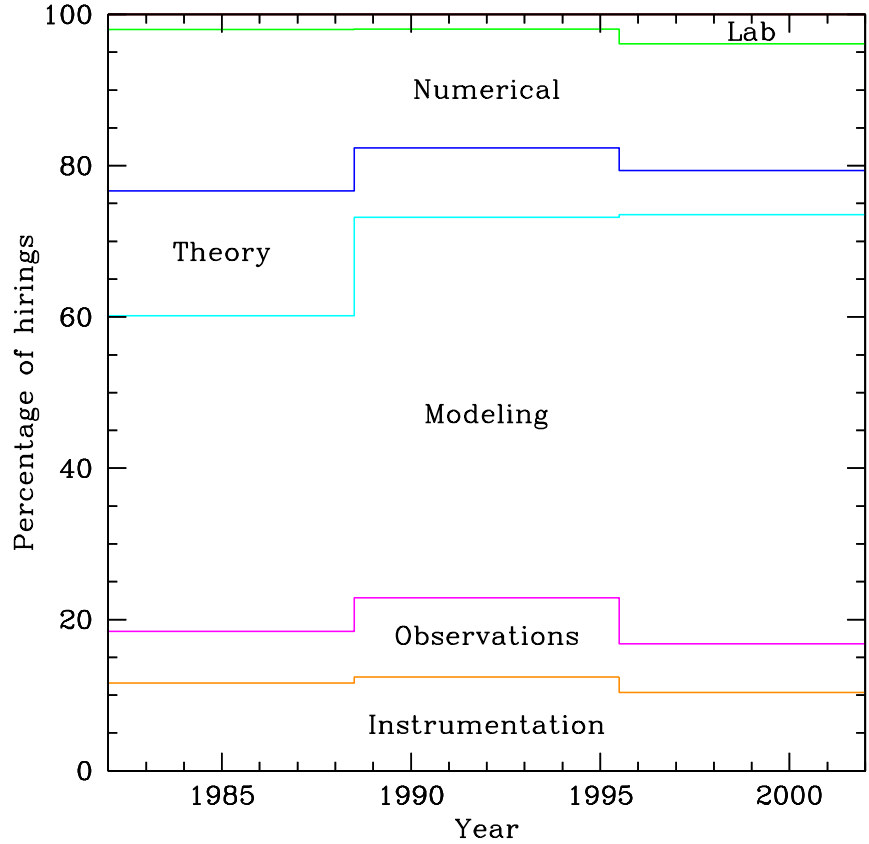,height=12cm}}
\caption{Time evolution of the methodologies of the new recruits in France.
>From {\em bottom} to {\em top}, the methodologies are:
Instrumentation, Observations, Modeling (which can include observations), 
Theory, Numerical, and Laboratory.}
\label{evolmeth}
\end{figure}

Fig.\ \ref{evolmeth} presents the evolution in the principal
methodologies of the
recruited astronomers and astrophysicists since 1982.

The trends are fairly stable, except for a statistically significant increase
in the fraction of scientists mainly involved in modeling, and a significant
decrease of scientists involved in theory.

\subsection{Fraction of women recruits}

In the last 21 years, 
the fraction of women recruits (in all four corps: CNRS, Observatories,
Universities and CEA) has remained essentially constant at 20\%:
20\% in 1982-1988, 22\% in 1989-1995, 19\% in 1996-2002 (the latter
decrease is not statistically significant).
However, the fraction of women hired in 1965-1975 was significantly higher,
at over 25\%.
The fraction of women hired by the four corps since 1980 are statistically
identical (19\%, 22\%, 20\% and 15\% for the CNRS, Observatories,
Universities and CEA, respectively).
The success rate of female candidates for entry-level 
CNRS positions is virtually the
same as their male counterparts (Durret 2002).

\subsection{Age of recruitment}
\label{ageevol}

\begin{figure}
\centerline{\psfig{file=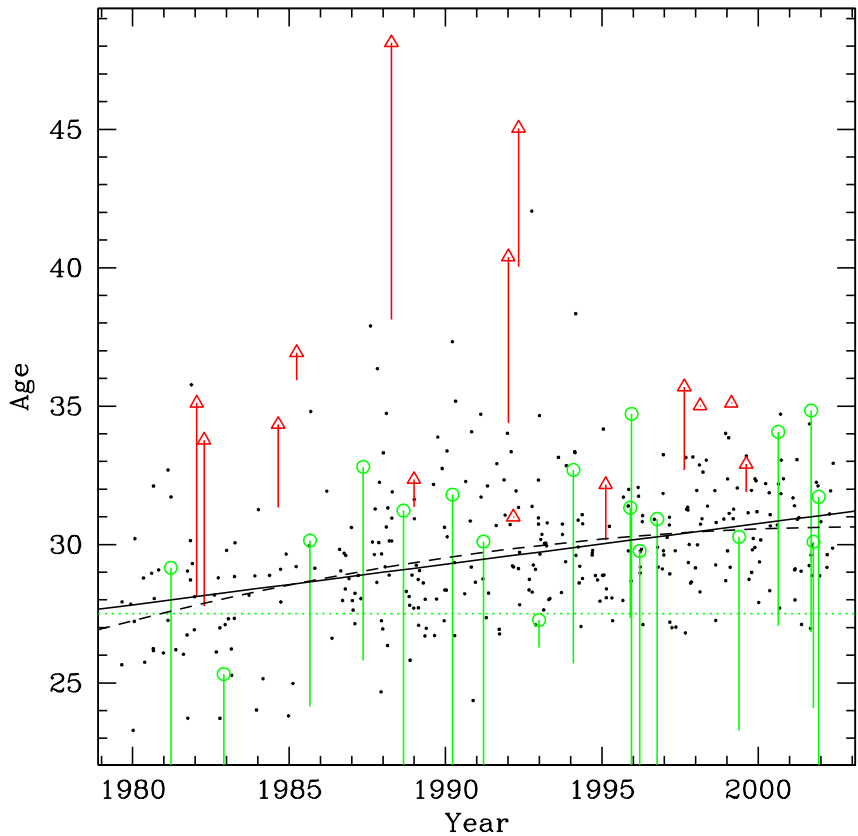,height=12cm}}
\caption{Age of scientists at the end of the year when they were hired. Only
CNRS, Observatory and University 
scientists hired to entry-level (including CNRS-CR1) positions are shown.
French scientists are shown as {\em small dots:}, scientists who came to
France before turning 28 are shown as {\em open circles}, while scientists
who came at 28 or later are shown as {\em triangles}.
Years and ages are slightly randomized ($\pm 0.4$ years) for clarity.
The {\em segments} going down from each scientist who came to France
indicate the ages when (s)he was already in France.
The {\em solid oblique line} shows the linear fit  (eq.\ [\ref{eqagevsyear}])
to the scientists 
always present in
France, with rejection of outliers.
The {\em dashed curve} is the analogous parabolic fit.
}
\label{evolage}
\end{figure}

Are French scientists recruited increasingly later in their career?
Fig. \ref{evolage} shows the age of recruitment (the age at the end of the
year when the recruitment is pronounced) for scientists in the major three
corps (CNRS, Observatories, and Universities -- no age data was available
for the CEA staff).

Fig.\ \ref{evolage} shows the rise with time in the age of the recruits,
with the relation (straight line in the Figure)
\begin{equation}
\rm Age = 27.8 + 0.15\, (Year - 1980) \ .
\label{eqagevsyear}
\end{equation}
This increase is statistically significant.
For example, subdividing the scientists into three 7-year 
periods: 1982-1988,
1989-1995, and 1996-2002, the distribution of recruitment ages is
statistically higher through the Kolmo\-gorov-Smirnov test from the 1st to the
2nd period (93.7\% confidence) and from the 2nd to the 3rd period (99.5\%
confidence). Also a Spearman rank test gives a probability of 
$3\!\times\!10^{-11}$ of such a rank correlation (0.36) by chance.
The parabolic fit suggests that this increase in the age of tenured
scientists is currently flattening.
In fact, the rank correlation test yields a significant rise in hiring age 
during 1980-1991, but an insignifcant rise since 1992.
The overall rise in recruitment age
is significant for both CNRS and
Observatory positions (whose age patterns are very similar), 
but not for University positions.

An increase of 1.5 years of age every 10 years constitutes a strong rise.
Since 1980, the typical recruitment age has thus risen 3 years, which
represents 10\% of the age of the scientist.
In 2002, the typical recruitment age (parabolic fit
in Fig.\ \ref{evolage}) is 30.6 years.
This corresponds to 3-4 years after the PhD.
In contrast, around 1980, scientists were typically hired by the end of their
PhD, and in the 60's and 70's
some scientists were granted tenured positions
before starting their PhD, even at the young age of 22!

\section{Comparison with other Countries}

In comparison with other countries, France can boast one of the highest rate
per capita
of hiring tenured scientists in astronomy and astrophysics.
The advantage is that this attracts highly qualified foreign scientists.
The disadvantage is that the quality threshold is perhaps lower than in
countries with few positions.

Another originality of the French system is its reliance on national hiring
committees (see following section).
In comparison with other countries, the assessment of candidates by experts
outside the committees (recommendation
letters) have little impact on committees (but they can have a great impact
on the candidate's two referees).

In the United States and many other countries, 
a short list is made, based upon the publication record,
the recommendation letters, the perceived dynamism and the academic
background, and only the candidates on the short list are interviewed both in
private and through an hour-long seminar.
In France,  the CNAP is forced, by rule of law, 
to conduct auditions on
{\em all the candidates}, which, even after splitting the committees in two
or three sub-jurys, and limiting the presentations of the candidates to 15 or
20 minutes, still requires a few days full time.

As in most systems, the candidate to the French system must play by the rules
of the game. In a system with national hiring, this includes not only making
strong contact with the laboratory where one wishes to work, but also
making presentations of one's
work across the country, 
because as is usually the case in any country, a candidate with a known
face has
more chances of being hired than an unknown face (the CNAP now has
photographs of the candidates and in general 
a fraction of the members have seen the
candidate during an audition or seminar).
A candidate to the Observatories must moreover develop a credible observational
service task with the staff at the laboratory where (s)he wishes to settle.

One drawback of the French system is that it is fixed by rules of law, hence
very rigid. This prevents the rapid hiring of outstanding foreign scientists 
({\em e.g.} during the demise of the former Soviet Union).
Nevertheless, a few outstanding foreign individuals have been hired 
(see Sect.\ \ref{foreign} below).

\section{Recurrent Questions}

\subsection{Local versus national hiring}

In most countries, the hiring of tenured faculty is conducted by the local
institutions. For French astronomy, roughly 1/3 of astronomers are
hired locally (through the Universities and the CEA), 
while the remainder are hired through the national CNRS and CNAP
committees. 
Having myself pursued my graduate studies and postdoc in the United States, I
first found this situation quite 
strange. In hindsight, I now believe that national hiring works better for
France. 

While local hiring has the advantage of developing a healthy
competition between the various laboratories, it has the drawback of leading
($\simeq$ 1/4 of the time) to the hiring of
less-qualified ``locals'' instead of more qualified ``exteriors''. 
In other words, some (probably a minority of) 
French scientists sitting on local tenure committees lack the
free-market spirit to strengthening science, but prefer instead to favor
former 
students, and national committees prevent such abuses.

\subsection{Big science versus creative individuals and the issue of
targeted positions}
\label{bigscience}

Another important question is how to maintain the best balance between the
wish to hire the brightest individuals and the need to supply manpower to the
large high-priority observational projects in which France is investing
heavily in equipment. In the last three years, 37\% of the CNRS positions
were {\em targeted} to either a particular field of astronomy (sometimes a
pluridisplinary field at the boundaries of astronomy), and/or a particular
international observatory 
(see Table 2).
The question of supplying manpower to the important observational programs is
also addressed in detail by the CNAP, which, since the mid '90s, has been
considering
service tasks on large observational projects as a prime and necessary
criterion for hiring (although there are no specifically
targeted Observatory positions).

The advantage of targeting positions for specific fields or projects is
obvious: to fill an important need at a given time.
Moreover, given persistent rumors, over the last dozen years, 
on the intentions of the French government
to merge the Observatories into the Universities, the aspect of service tasks
performed by Observatory astronomers is considered to be the prime motivation
for keeping a specific status for
Observatory positions.
However, in most cases, the pressure on the targeted positions is
considerably lower than on the non-targeted ones:
Table 2 indicates that none of the CNRS targeted positions since 2000 had
pressures above 40\% of those of their non-targeted counterparts of the same
year, and most had a pressure 1/4 less than for Observatory positions.
One is therefore led to
wonder whether the targeted positions are attributed on average to
individuals of lower competence than are the non-targeted ones.

In the author's opinion, the pendulum is swinging a little 
too far in the direction of
supplying the big science projects, given that all the Observatory positions
(through the observational service task)
and nearly 40\% 
-- see Table 2 --
of recent CNRS positions are targeted to projects, and moreover,
over half of the University positions come with narrowly defined targets.
What is worse, the French
government recently proposed (but then withdrew in face of
the clear opposition of the astronomical community) a reform of the CNAP that
would have made all Observatory 
positions targeted both by field and by location (to
conform with the targeting of university positions).
It would make more sense to stop targeting CNRS positions, except in rare
instances for
specific interdisciplinary fields (the CNRS has partially solved this problem
by just creating a special committee
separate from {\em Solar System and Distant Universe} called
{\em Astroparticles}).
To provide manpower to the large projects, the INSU should release
an official
{\em public list}
of top-priority observational programs, that would contain
roughly twice the number of annual Observatory positions available, and
that would be renewed every year. The CNAP would then be required to hire
Astronomers proposing service tasks with one of these listed
programs.
With such a system, the CNRS would hire the more creative individuals, while
the CNAP would hire the team players. Both sets of scientists are necessary
to make a country like France thrive in the international arena.

\subsection{The hiring of foreign scientists}
\label{foreign}

Overall, since 1980, roughly
11\% of all scientists hired at the entry level were from foreign
countries or French nationals who (like this author) had left France before
graduate school, and both are hereafter refererred to as ``foreign''.
Table 3
shows the  numbers and fractions of foreign scientists
per corps (these are lower limits, as some may have been missed).

\begin{table}[ht]
\begin{center}
\caption{Foreign scientists (including French expatriates) hired since 1980,
per corps}
\begin{tabular}{lr@{\ \ \ \ }r@{\ \ \ \ \ \ \ \ \ }r@{\ \ \ \ \ \ \ \ }r@{\ \
\ \,}r@{\ \ \ \ }}
\hline
        & \multicolumn{1}{c}{CNRS} & \multicolumn{1}{c}{Observatories} &
        \multicolumn{1}{c}{Universities} & \multicolumn{1}{c}{CEA} &
        \multicolumn{1}{c}{Total} \\ 
\hline
Foreign & 21 & 15 & 2 & 7 & 45 \\
Total   & 169 & 121 & 79 & 33 & 402 \\
Percentage & \multicolumn{1}{r@{\%\ \,}}{12} 
& \multicolumn{1}{r@{\%\ \ \ \ \ \ \,}}{12} 
& \multicolumn{1}{r@{\%\ \ \ \ \ \,}}{3} 
& \multicolumn{1}{r@{\%\ }}{21} 
& \multicolumn{1}{r@{\%\ \,}}{11} \\
\hline
\end{tabular}
\end{center}
\label{foreigntb}
\end{table}

The Table indicates a statistically significant low fraction of foreign
scientists hired at the Universities (given a predicted foreign-hiring 
rate of 11\%, there is 0.6\% probability that two or less foreigners
would be hired out of 79 University recruits).
This low fraction of foreign University recruits
presumably arises because 
foreign accents are
deemed a hindrance to good teaching, and also because foreign scientists are
often aware too late of the prerequisite of qualification by the CNU.
On the other hand, the CEA has a strong tradition of hiring foreign
scientists (21\% of their hirings since 1980, which is 
statistically significantly
high).

The origins of the foreign recruits are Western Europe (58\%),
South America (10\%), 
North America (9\%), 
North Africa (7\%), 
Middle East (7\%),
Former Soviet Union and Eastern Europe (5\%), 
East Asia (2\%),
and Australia (2\%).

Fig.\ 4
helps answer the question of the openness of the French
system to scientists from outside France.
Overall, the median delay for the hiring of foreign scientists, relative
to their French
counterparts (excluding those older scientists directly hired at upper-level
positions) is 2.9 years.
The age of arrival in France makes a crucial difference.
Scientists arriving before the age of 28 are hired 1.0 year (median) 
after their
French counterparts, while scientists arriving at 28 or after are hired 5.5
years (median) after their French counterparts.
Only three (late arrivals) out of 31
foreign scientists (10\%)
were hired directly
from abroad.
In addition to that, 7 foreign scientists were directly hired (in the last 23
years) to upper
level positions, among which 2 directly from abroad (both for positions of
head of their laboratory), and one only one year after his arrival
(again for a position of head of a laboratory).

Since the time of arrival is essential, one can wonder whether late-arrivals
of foreign scientists lead to longer or shorter time intervals to tenure in
comparison with earlier arrivals.

\begin{figure}
\centerline{\psfig{file=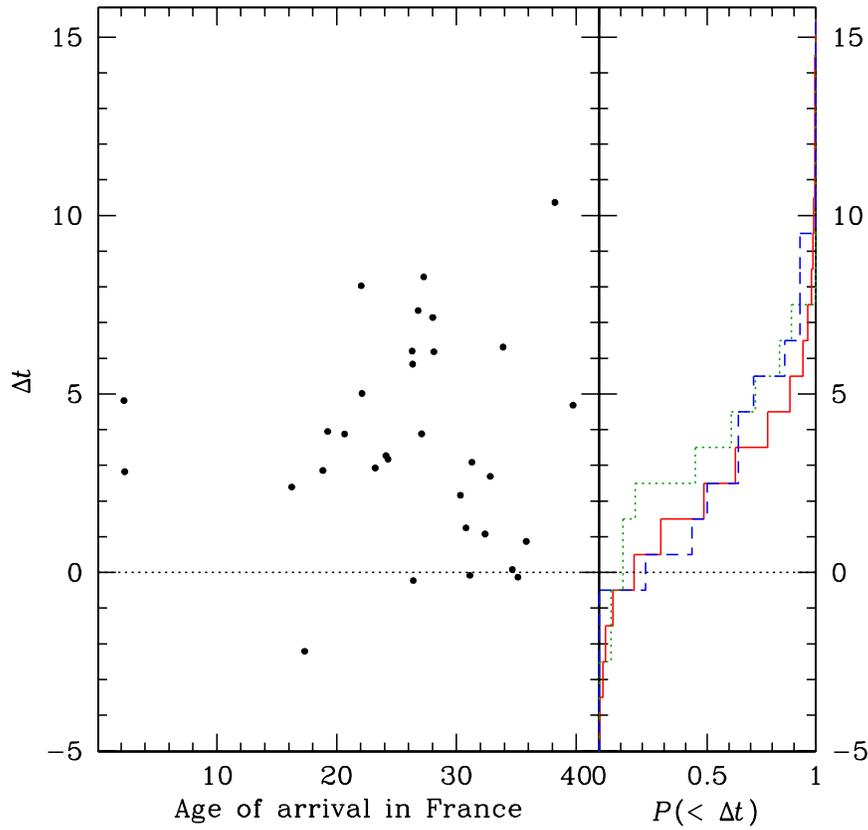,height=11.8cm}}
\caption{{\em Left:}
Time interval for hiring, measured from time of arrival in France or
the age of 27, whichever comes later, as a function of
the age of arrival for foreign astronomers or French expatriates.
{\em Right:} Normalized cumulative histograms
of time interval of hiring (defined above) for French nationals ({\em solid
histogram}), foreign scientists arriving before ({\em dotted
histogram}) and after ({\em dashed histogram}) the age of 27.}
\label{evolage2}
\end{figure}

The plot on the left of Fig.\ \ref{evolage2} shows no correlation between
the age of arrival and the
time interval for hiring measured from the age of arrival
if greater than 27 (corresponding approximately
to the year of the PhD thesis), or else from 27 years of age.
With this definition of the time interval,
foreign scientists wait globally
0.8 year more (mean) than their French counterparts
for securing tenured positions (this difference is significant to 94\%
confidence). In fact, the plot on the right of
Fig.\ \ref{evolage2} shows the 1 year delay for foreigners arriving early
(99\% significant with a KS test)
with only 17\% hired by the age of 29, in comparison with the 48\% of their
French counterparts.
On the other hand, foreigners arriving after 27
typically wait 2.5-3 years, slightly less (but not significantly so)
than it takes French nationals
counting from the age of 27.

The penalty on the foreign astronomers who arrive young in France may be due
to cultural differences. This is even more pronounced for foreign scientists 
arriving later in France, but although they
are hired at older ages, they wait the same (once in France) then do their
French counterparts (now counting from the age of 27).

\subsection{How important should be the criterion of 
integration within a laboratory?}

\parskip=0pt
In France, the perceived integration of the candidate to the laboratory where
he wishes to work is considered an essential selection criterion.
The heads of laboratories are asked by the CNAP (but not by the CNRS
committee) to present an
ordered list of their preferred candidates, and this information is very
influential: in the last two
years, among the 14 scientists that the CNAP has hired at entry-level
positions, only one had not been at
the top of the list by the laboratory he was applying to work at.

The advantage of such a system is that it enables the heads of laboratories,
often with the help of the scientists working in their lab,
to plan the recruitments and allows them to strengthen (or weaken) particular
research teams within their laboratory.
The drawback is that, just like in the case of targeted positions, the
overall quality of hired scientists will not be optimal.
For example,
the most brilliant candidates may not be high-priority in their laboratories,
either because their subfield of work is not well 
represented in their laboratory,
or because they are applying to an outside laboratory (not one where they
performed their doctoral thesis work) and the laboratory head is prioritizing
one or more candidates from his/her laboratory.
This is a likely occurrence when excellent foreign scientists seek a
laboratory where to apply for tenure.

In this author's opinion, a brilliant scientist who settles in a laboratory
with nobody else in his/her particular subfield of work, will manage, in
time, to 
attract  others in his/her subfield, some of whom may be former students.
It is not important if, during his/her first years of tenure, 
such a scientist does not publish
with other scientists in his laboratory. In this age of electronic mail and
cheap telephone rates, long-distance collaborations are easy to implement.
The long-term benefits outweigh the lack of early collaborations.

\subsection{Is recruitment governed by chance?}

\parskip=0pt
One would certainly hope that chance is absent in the recruitment
procedures of French astronomers.
Unfortunately, there is indeed an element of chance in the French tenure
system, as there is on any tenure (or telescope allocation) committee.
Indeed, the forcefulness of the candidate's referees is an important element
of the decision making in any system that does not rely {\em too strongly} on
bibliometric quantities such as the rates of paper production and/or citations.

Another element that could be considered chance is
the competition that the candidate might suffer in a given laboratory and/or
field of work. However, if the candidate feels overshadowed by another
candidate to the same laboratory, (s)he may well apply to a different one.

Finally, in the competition for positions at the Observatories, the candidate
must make the right choice of an observational service task that is in a
large observational program prioritized by the national funding agency
(INSU). If an official list of such prioritized programs (as advocated at the
end of Sec.\ \ref{bigscience})
were
publicly made
available roughly half a year before the season of tenure committees (usually
Spring
in France), candidates would then 
be able to begin work on a publicly listed observational service
task before launching their candidacy.

One can ask why did some of the best candidates fail to obtain a tenured
position in France after many tries.
The best known few cases involved scientists (who usually
find permanent astronomical positions
outside the country)
with important gaps in their
publication record.
Although the committees in France avoid basing their decisions by simply
counting papers, they are often
strongly tempted to pass on a candidate with a large gap in his/her publication
record, in 
favor of other candidates of similar overall perceived competence
with large publication records, and hence considerably more numerous
noteworthy scientific results.

\subsection{How much mobility should be allowed between laboratories
and between corps?}

Another question concerning the human resources of a given country, is
whether the decision makers should strive to strengthen the strongest
laboratories and abandon the weakest, or on the contrary 
provide a minimum human workforce for all laboratories.

This question affects the decisions of the national tenure committees (CNRS
and CNAP), and the second view point is probably at least as well followed as
the first.
This question also affects the rules regarding the mobility of French
scientists. Some believe that scientists should be free to change
laboratories, if they are accepted by a new one. Others believe that 
mobility should be made more difficult, to prevent
numerous departures from a given laboratory.

In this author's opinion, the first view point is to be favored, as leading to
optimal scientific productivity. If a scientist feels frustrated enough to
want to leave his/her laboratory, which entails a loss of a few weeks in
negotiations, moving and settling in the new laboratory and location, this
must mean that (s)he senses a considerable increase in his/her scientific
productivity in the new laboratory. Moreover, some scientists will want to
move for family reasons.

If the situation in a given laboratory is 
serious enough to lead to many scientists wanting to leave, should they be
forced to stay? Would all the good scientists flock to the most reputed
laboratories? Some would indeed, 
but only up to some point, when the laboratories will
have filled their office space. 
And the national tenure committees (CNRS and CNAP) can attempt to prioritize
the hiring of scientists in the laboratories that have recently lost
scientists, which makes more sense if this loss is caused by premature death
than by retirement, or especially voluntary leave.

\section{Summary and recommendations}

The French tenure system is probably unique in the world, with a high rate of
hiring per capita,
its combination
of local and national hiring tenure committees, the young age at which its
scientists reach tenured positions, and its governance by state rules and
decrees.

The balance between the free-market and organized/planned approaches is also
unique to 
France. 
The strong reliance on national tenure committees is justified as
preventing abuses of favoring local candidates.
On the other hand, although
the increased targeting of tenured positions by subfield and/or laboratory
is obviously helping France provide manpower in its large priority projects,
it is also leading to a corresponding decrease in the
hiring of the brilliant and creative scientists.
This situation can be remedied by a {\em public list} of $2\,N$ priority
projects set by INSU from which the CNAP committee will select the service
tasks of its $N$ Observatory positions,
and a reduction by at least half
of the fraction of targeted CNRS positions.
Moreover, one should be careful not to prevent the mobility of scientists in
the name of laboratory planning.

The French system, although rigid by virtue of its being governed
by government decrees and ruling, is nevertheless flexible enough to hire a
substantial fraction (11\%) of foreign scientists.  
Young arrivals in France are hired typically one year after their French
counterparts, while those arriving after their PhD typically wait 2.5-3 years
before being given tenure.

Still, astronomers and astrophysicists working in
French laboratories are at least as influential
as scientists working in other Western countries (Bertout {\em et al.} 2003).
In some respects, 
the French system is slowly converging to the other western systems, in that
most young scientists now go for postdoctoral positions outside the country,
and, in general, now publish almost all their
works in English, are more present in international
conferences and committees, and are now more experienced (typically age 31)
when they are granted
tenured positions.
Nevertheless, {\em vive la diff\'erence!}

\section*{Acknowledgements}

I warmly thank 
J.C. Vial and A. Heck for useful comments, and
F. Durret, 
B. Guiderdoni, 
R. Ferlet, 
M. Marcelin, 
R. Mochkovitch,
and many others for useful general or detailed information for the
statistical study presented here.

\end{document}